# Strategic AI adoption in SMEs: A Prescriptive Framework

**Atif Hussain**
atif.iqtm@pu.edu.pk
University of the Punjab

**Rana Rizwan**
rana.rizwan@solutyics.com
Solutyics (Private) Limited

**Abstract**

Artificial Intelligence (AI) is increasingly acknowledged as a vital component for the advancement and competitiveness of modern organizations, including small and medium enterprises (SMEs). However, the adoption of AI technologies in SMEs faces significant barriers, primarily related to cost, lack of technical skills, and employee acceptance. This study proposes a comprehensive, phased framework designed to facilitate the effective adoption of AI in SMEs by systematically addressing these barriers. The framework begins with raising awareness and securing commitment from leadership, followed by the adoption of low-cost, general-purpose AI tools to build technical competence and foster a positive attitude towards AI. As familiarity with AI technologies increases, the framework advocates for the integration of task-specific AI tools to enhance efficiency and productivity. Subsequently, it guides organizations towards the in-house development of generative AI tools, providing greater customization and control. Finally, the framework addresses the development of discriminative AI models to meet highly specific and precision-oriented tasks. By providing a structured and incremental approach, this framework ensures that SMEs can navigate the complexities of AI integration effectively, driving innovation, efficiency, and competitive advantage. This study contributes to the field by offering a practical, prescriptive framework tailored to the unique needs of SMEs, facilitating the successful adoption of AI technologies and positioning these organizations for sustained growth in a competitive landscape.

## 1. INTRODUCTION

Artificial Intelligence (AI) is increasingly recognized as a crucial element for the advancement and competitiveness of modern organizations (Kordon, 2020; Prasad Agrawal, 2023). Despite its potential, the adoption of AI technologies faces significant barriers that hinder its implementation, particularly in Small and Medium Enterprise (SMEs). These barriers include cost, lack of technical skills, and employee acceptance (Auer et al., 2023; Oldemeyer et al., 2024; Ulrich & Frank, 2021).

Existing studies on adoption of AI in organizations have mainly relied on established models like Technology-Organization-Environment (TOE) (Polisetty et al., 2024), Technology Acceptance Model (TAM) (Rahman et al., 2023) or People-Process-Technology Model (Uren & Edwards, 2023). While these models provide valuable explanatory insights into the factors influencing AI adoption, they fall short in offering practical, actionable guidelines that managers can directly apply in their organizational contexts. These models are primarily theoretical and do not sufficiently address the specific challenges faced by SMEs, which often have limited resources and unique operational constraints.

Given these limitations, there is a pressing need for a prescriptive model that not only takes into account the barriers to AI adoption but also provides clear, actionable guidelines for overcoming these obstacles. Such a model would be particularly beneficial for SMEs, which stand to gain significantly from AI but are often hindered by the aforementioned barriers.

Accordingly, we propose a prescriptive framework for AI adoption in SMEs. This framework is designed to be practical and actionable, providing managers with clear steps and strategies for implementing AI in their organizations. In developing this framework, we have set the following goals:

1. The framework should be straightforward and easy to comprehend, ensuring that managers can easily follow it.
2. The framework should specifically address the common barriers to AI adoption in SMEs, such as cost, lack of technical skills, and employee resistance, providing strategies for overcoming these challenges.
3. Unlike existing explanatory models, the framework should offer concrete, actionable guidelines that managers can follow to successfully adopt AI technologies.

The rest of the article is structured as follows. In section 2, we describe methodology, followed by the description of the proposed framework in section 3. Discussion on how the proposed framework meets the objectives specified in the introduction is presented in section 5. Finally, section 6 provides a conclusion.

## 2. METHODOLOGY

We have employed a reflection and synthesis approach to develop a prescriptive framework for AI adoption in SMEs. This methodology involves integrating existing fragments of insights from various sources, including theoretical models, empirical studies, and practical experiences (Jaakkola, 2020). By reflecting on these diverse perspectives, we aim to construct a cohesive and practical guide tailored to the unique needs and constraints of SMEs. We draw on a wide array of existing knowledge and experiences to inform our framework. This approach allows us to capture the multifaceted nature of AI adoption, considering both the theoretical underpinnings and the practical challenges faced by SMEs. By synthesizing these insights, we aim to provide actionable strategies that address the specific barriers SMEs encounter, such as cost, lack of technical skills, and employee acceptance, thus facilitating effective AI implementation

## 3. THE FRAMEWORK

This section outlines the structured, phased framework designed to facilitate the adoption of AI in small and medium enterprises (SMEs). The proposed framework comprises the following phases:

1. Awareness for Leadership
2. Adoption of General-Purpose Generative AI Tools
3. Adoption of Off-the-Shelf Task-Specific Tools
4. In-House Development of Generative AI Tools
5. In-House Development of Discriminative AI Models

Each phase includes specific activities, policy initiatives, review processes, and associated costs, providing a comprehensive guide for SMEs to navigate the complexities of AI adoption effectively which is summarized in Table 1. Each of these phases is discussed next.

**3.1 Awareness for Leadership**

Management support is often a key enabler of successful adoption of an initiative in a company (Hsu et al., 2018; Kurup & Gupta, 2022). Therefore, the first step should be to educate and engage leadership on the potential and impact of artificial intelligence (AI). Cultivating a comprehensive understanding among leadership can ensure strategic alignment and secure commitment to AI initiatives.

To achieve this objective, workshops and seminars for leadership teams should be organized. These can play a crucial role, as awareness of technology significantly drives its adoption (Subedi et al., 2009). Such sessions should provide an extensive overview of AI technologies, covering the basics, current trends, and future possibilities. The aim should be to demystify AI for executives to help them understand what it has to offer.

Sharing success stories and case studies from similar industries can help demonstrate the practical applications and benefits of AI. These real-world examples enable leaders to visualize how AI can be effectively applied within their own organization. Moreover, case studies and success stories can create memetic isomorphic pressure, fostering a desire to adopt the technology (Lai et al., 2006; Ukobitz & Faullant, 2022).

Post-workshop feedback should be collected from leadership teams to assess understanding and address any concerns, refining future sessions to ensure that leadership remains well-informed and aligned with AI initiatives.

In terms of policy formulation, an initial AI ethics and governance policy should be developed, addressing ethical considerations and governance structures for AI use. This policy should establish foundational principles for AI adoption within the organization, prioritizing ethical considerations from the outset.

## 3.2 Adoption of General-Purpose Generative AI Tools

The current interest in AI has been fueled by the release of generative AI tools. Examples include OpenAI's ChatGPT, Google's Gemini (erstwhile Bard), and Microsoft's Copilot (Szczesniewski et al., 2024). These are general purpose tools which can be used for a variety of tasks in a number of domains (Aguinis et al., 2024; Biswas, 2023; Firat, 2023; George & George, 2023; Rane, 2023).

These general-purpose tools mostly have both the free and paid versions and thus serve as a good starting point for a company looking to adopt AI as employees can be encouraged to start using these tools without any cost commitment. Employees should be given access to these general-purpose tools asked to integrate them into daily operations. This allows them to interact with and utilize AI technology practically. To facilitate this introduction, training sessions and hands-on workshops should be organized to ensure employees are comfortable using these new tools. Training increases individuals' knowledge which in turn creates a positive attitude towards a new technology (Mullins & Cronan, 2021). These sessions should cover the functionalities and benefits of the AI tools, providing practical demonstrations and opportunities for practice. Particularly, training for prompt engineering (Ekin, 2023) should be provided. Increased use and experimentation with a technology fosters a positive attitude in individuals (Tubaishat et al., 2016).

Regular user feedback should be collected to identify areas for improvement for smooth integration of these tools into workflows. Additionally, usage analysis reviews should be conducted regularly to measure the impact and effectiveness of the AI tools, identifying any issues or areas for enhancement.

In terms of policy formulation, data privacy and usage policies should be developed, outlining how data will be used, stored, and protected when interacting with AI tools. This includes ensuring compliance with legal standards and maintaining user privacy and data security.

### 3.3 Adoption of Off-the-Shelf Task-Specific Tools

Once the employees have become familiar with general purpose tools and have experienced the power of generative AI, the next step is to use task specific tools. Many established companies and startups are developing task specific tools on top of generative AI models which excel at doing a specific task (*Future Tools - Find The Exact AI Tool For Your Needs*, n.d.; *There's An AI For That (TAAFT) - The #1 AI Aggregator*, n.d.). These tools allow enhancing efficiency and productivity in specific areas. To decide which tools to use, a comprehensive assessment should be conducted to identify key tasks and processes that can benefit from AI optimization. This involves analyzing current workflows and pinpointing areas where AI can add significant value. Based on the assessment, suitable off-the-shelf AI solutions should be selected and implemented. These solutions must be chosen for their ability to address specific organizational needs (Alami et al.,

2020). Comprehensive training should be provided to ensure employees are proficient in using the new AI tools. This includes integrating the tools into existing workflows and offering ongoing support to address any challenges.

It must be noted here that unlike general purpose tools, these tools are mostly paid, with free versions offering only limited functionality. However, since these tools are available as SaaS (Tsai et al., 2014), no upfront or infrastructure cost are required and companies can start using these tools for minimal subscription costs depending on their usage.

In terms of policy formulation, task-specific AI policies should be developed to ensure effective and ethical use of AI tools in various contexts. Further, since AI initiatives must remain aligned with organizational strategy (Kitsios & Kamariotou, 2021), regular reviews are also essential to assess the effectiveness of AI tools in meeting organizational goals, focusing on efficiency improvements and overall productivity gains. Additionally, gathering feedback from employees on their experiences and challenges with the new tools will facilitate necessary adjustments and improvements.

### 3.4 In-House Development of Generative AI Tools

In stage 2 and 3, the focus is on using readily available tools; however, once organizations have gained experience with off-the-shelf products and become more familiar with AI and its potential, they may start to see new opportunities or better ways to use AI that these standard solutions cannot support. This growing familiarity and understanding of AI capabilities can highlight the limitations of off-the-shelf products, prompting organizations to consider developing in-house generative AI tools.

Generative AI models can be easily leveraged for building new applications (Weber, 2024), particularly with the availability of frameworks created specifically for this purpose (Pinheiro et al., 2023; Topsakal & Akinci, 2023). By doing so, organizations can achieve greater customization, control, and integration with their specific business needs. In-house development also enhances data privacy and security, offers potential long-term cost efficiencies, and provides the flexibility to innovate and adapt quickly to evolving market conditions or internal requirements. Therefore, while readily available tools are valuable for initial stages, deeper engagement with AI often drives the shift toward bespoke AI solutions to fully leverage the technology's potential.

At this point, organizations have the option to build products on top of available proprietary or open-source models. They can either hire a dedicated team to develop these systems in-house or hire a few individuals and outsource the development.

When choosing between proprietary and open-source models, organizations must consider their specific needs and usage patterns. Proprietary models can be cost-effective if the usage is low, as they often come with lower initial costs and bundled support. However, open-source models offer more control and flexibility, which can be crucial for customization and innovation. The downside of open-source models is the potential for higher costs due to cloud infrastructure requirements. Nevertheless, if the usage is high, these costs can decrease, making open-source models a more viable option in the long run.

Establishing an AI development team or partnering with external developers to create tools aligned with organizational requirements is needed at this stage. The team, whether internal or external, should collaborate with various departments to identify use cases that add significant value, guiding the development process to address real organizational needs. The development team will design, test, and deploy these tools, ensuring they meet performance and reliability standards before integrating them into workflows. Continuous training and support should be provided to ensure employees are proficient in using the tools, maximizing their benefits.

Policy formulation in this phase should include creating comprehensive guidelines for the development and deployment of AI tools, ensuring ethical, secure, and effective processes. These policies cover coding standards, testing protocols, and deployment procedures. Regular reviews of development milestones will help track progress and ensure the development stays on course. User feedback should also be continuously collected and incorporated to refine and improve the AI tools, ensuring alignment with user needs and organizational objectives.

**Phase 5: In-House Development of Discriminative AI Models**

This phase focuses on the development of discriminative models, which are essential for tasks requiring high precision and specificity. Discriminative models, unlike generative models, are designed to differentiate between classes and make accurate predictions based on input data. These models are crucial for applications such as fraud detection, customer segmentation, and predictive maintenance, where accuracy and reliability are paramount.

Unlike previous phases, where organizations could leverage existing tools and build on top of generative AI models, this phase involves developing discriminative models from scratch. This process requires a significant investment in infrastructure, data collection, model training, and the establishment of dedicated teams and policies. The complexity and specificity of these models demand a more comprehensive approach to ensure they meet the precise needs of the organization.

The first step in this process involves building the necessary infrastructure to support the development and deployment of discriminative models. Organizations need to decide between cloud-based solutions and on-premises data centers. Cloud solutions such as AWS, Google Cloud, and Azure offer scalability and flexibility, making them suitable for organizations looking to quickly scale their operations. However, on-premises solutions might be preferable for organizations dealing with highly sensitive data or seeking long-term cost efficiency. Investing in high-performance computing resources, such as GPUs and TPUs, is essential to handle the computational demands of training complex models. Additionally, robust data storage solutions, including data lakes and data warehouses, should be implemented to efficiently manage and retrieve large volumes of data.

Effective data collection and management are critical for training discriminative models. Organizations need to identify and integrate diverse data sources relevant to their specific use cases, such as transactional data, customer interactions, and sensor data. Establishing a data governance framework ensures data quality, consistency, and security across the organization. This includes setting up protocols for data cleaning, preprocessing, and augmentation to prepare datasets for training models. Adopting advanced data management tools and techniques, such as automated data pipelines and real-time data processing, can enhance efficiency and accuracy in data handling.

Training and optimizing discriminative models require meticulous planning and execution. This involves selecting appropriate algorithms and architectures tailored to specific tasks. Implementing an iterative process of model development, testing, and validation ensures that models achieve high performance and generalization capabilities. Utilizing hyperparameter tuning and techniques like cross-validation can further refine model accuracy. Maintaining a comprehensive record of experiments and results aids in tracking progress and making informed adjustments.

Having a skilled team is paramount to the success of discriminative model development. Recruiting data scientists, machine learning engineers, and domain experts with the necessary expertise in AI and machine learning is crucial. Continuous training and professional development opportunities should be provided to keep the team updated with the latest advancements in the field. Establishing a collaborative environment where team members can share knowledge and insights fosters innovation and problem-solving.

Policy formulation at this stage plays a critical role in ensuring ethical, secure, and effective use of discriminative models. Developing comprehensive guidelines that cover aspects such as data privacy, ethical AI use, and compliance with legal standards is essential. These policies should also address coding standards, testing protocols, and deployment procedures to ensure the reliability and security of AI systems. Regular reviews and updates of policies are necessary to adapt to evolving technological and regulatory landscapes. Collecting user feedback continuously and incorporating it into policy adjustments ensures alignment with organizational goals and user needs.

Developing discriminative models from scratch involves significant costs, including investments in infrastructure, data collection, staff recruitment, and ongoing training. However, the long-term benefits of having highly accurate and customized AI solutions can outweigh these initial expenses. Organizations should conduct a thorough cost-benefit analysis to ensure that the investments align with their strategic objectives and provide a clear return on investment.

| Table 1: Summary of AI Adoption Phases | | | | |
|---|---|---|---|---|
| Phase | Main Activities | Policy Initiatives | Reviews | Associated Costs |
| Phase 1: Awareness for Leadership | Organize workshops and seminars for leadership; Provide extensive overview of AI technologies; Share success stories and case studies from similar industries; Collect post-workshop feedback | Develop initial AI ethics and governance policy | Assess understanding and address concerns; Refine future sessions | Cost of organizing workshops and seminars; Potential consulting fees |

| Phase 2: Adoption of General Purpose Generative AI Tools | Provide access to general-purpose AI tools; Organize training sessions and hands-on workshops; Train for prompt engineering; Collect regular user feedback; Conduct usage analysis reviews | Develop data privacy and usage policies | Identify areas for improvement; Measure impact and effectiveness | Subscription costs for AI tools (if applicable); Training program costs |
|---|---|---|---|---|
| Phase 3: Adoption of Off-the-Shelf Task-Specific Tools | Conduct assessment to identify key tasks for AI optimization; Select and implement suitable off-the-shelf AI solutions; Provide comprehensive training; Collect employee feedback | Develop task-specific AI policies | Regular reviews to assess tool effectiveness; Gather employee feedback | Subscription costs for task-specific tools; Training program costs |
| Phase 4: In-House Development of Generative AI Tools | Establish AI development team or partner with external developers; Identify valuable use cases; Design, test, and deploy tools; Provide continuous training and support; Collect user feedback continuously | Create guidelines for development and deployment of AI tools; Ensure ethical, secure processes | Regular review of development milestones; Incorporate user feedback | Salaries for in-house developers or outsourcing fees; Infrastructure costs |
| Phase 5: In-House Development of Discriminative AI Models | Build infrastructure for model development; Establish data collection and management protocols; Select and optimize appropriate algorithms and architectures; Provide continuous training and support; Conduct rigorous testing and validation | Develop comprehensive guidelines for data privacy, ethical AI use, coding standards, and deployment procedures | Regular reviews and updates of policies; Continuous user feedback | Significant investment in infrastructure; Salaries for specialized staff; Data management and storage costs |

## 4. DISCUSSION

The proposed framework methodically addresses and mitigates the primary barriers to AI adoption in SMEs: cost, lack of technical skills, and employee acceptance. Through a structured and phased approach, this framework ensures SMEs can effectively navigate the complexities of AI integration and leverage its potential.

### 4.1 Addressing Cost Barriers

One of the predominant challenges for SMEs in adopting AI is the significant initial investment required. This framework mitigates this concern by recommending the initial use of general-purpose generative AI tools, many of which are available in free or low-cost versions. This allows organizations to experiment with AI technologies without incurring substantial financial commitments. As organizational familiarity and comfort with AI technologies increase, the transition to more specialized, task-specific tools, and eventually to in-house development, can be justified by demonstrated returns on investment and operational improvements. This phased investment strategy not only spreads costs over time but also aligns expenditures with the organization's growing capabilities and confidence in AI technologies.

**4.2 Enhancing Technical Competence**

The lack of technical expertise is a significant barrier to AI adoption. The framework systematically builds technical skills within the organization through continuous training and hands-on workshops integrated at each stage. Beginning with basic awareness and progressing to advanced technical training, the framework ensures that employees gradually acquire the requisite skills to operate and innovate with AI tools. By initially focusing on accessible and user-friendly AI applications, employees can build a foundation of knowledge and confidence, which is then expanded through targeted training programs for more complex, task-specific, and in-house developed AI solutions. This incremental learning process effectively mitigates the skill gap and fosters a culture of continuous learning and technological fluency within the organization.

**4.3 Fostering Employee Acceptance**

Employee resistance to new technologies can significantly impede AI adoption. The framework addresses this issue by involving employees early in the adoption process and providing extensive training and support. Workshops and seminars aimed at both leadership and general staff help demystify AI, illustrating its potential benefits and practical applications through real-world case studies. Regular feedback mechanisms ensure that employee concerns are heard and addressed, thereby enhancing buy-in and reducing resistance. By actively engaging employees and demonstrating the tangible benefits of AI through improved workflows and efficiencies, the framework cultivates a positive attitude toward AI technologies.

**4.4 Establishing Comprehensive Policy and Governance**

Effective AI adoption necessitates robust governance frameworks to address ethical, legal, and operational concerns. The framework emphasizes the development of comprehensive policies at each stage, including AI ethics, data privacy, and usage guidelines. These policies ensure that AI implementations are not only effective but also aligned with organizational values and regulatory requirements. By establishing clear governance structures and ethical guidelines from the outset, the framework builds trust and accountability, both within the organization and externally with stakeholders.

**4.5 Promoting an Incremental and Scalable Approach**

The phased approach of the framework allows SMEs to adopt AI technologies incrementally, starting with general-purpose tools and progressing to more advanced, customized solutions. This scalability ensures that organizations can adapt AI initiatives to their unique contexts and resource constraints, scaling up as they grow in capability and confidence. Each phase builds on the previous one, creating a solid foundation for more sophisticated AI applications and thereby reducing the risk of failure and ensuring sustained progress.

## 5. CONCLUSION

This article presents a comprehensive framework for the adoption of artificial intelligence (AI) in small and medium enterprises (SMEs), addressing the primary barriers of cost, lack of technical skills, and employee acceptance. By employing a phased, incremental approach, the framework ensures that AI integration is both practical and effective, providing a clear pathway for SMEs to leverage AI technologies for competitive advantage.

The proposed framework begins with raising awareness and securing commitment from leadership, a crucial step for any successful technology adoption. By organizing workshops and seminars, and sharing success stories and case studies, leadership teams are better equipped to understand the potential and impact of AI. This initial phase ensures strategic alignment and prepares the organization for the subsequent stages of AI integration.

The second phase involves the adoption of general-purpose generative AI tools, which offer a low-cost entry point for SMEs. These tools allow employees to interact with AI technologies in a practical manner, facilitating hands-on learning and experimentation without significant financial investment. This phase not only builds technical competence but also fosters a positive attitude towards AI among employees.

As the organization becomes more comfortable with AI technologies, the third phase involves the adoption of task-specific AI tools. This targeted approach enhances efficiency and productivity in specific areas, providing clear examples of AI's value within the organization. Comprehensive training and regular feedback mechanisms ensure that employees can effectively integrate these tools into their workflows.

The fourth phase focuses on the in-house development of generative AI tools. This stage provides greater customization and control, addressing specific business needs and enhancing data privacy and security. Establishing an AI development team or partnering with external developers allows the organization to leverage advanced AI capabilities while ensuring ethical and secure processes.

Finally, the fifth phase involves the development of discriminative AI models from scratch, addressing highly specific and precision-oriented tasks. This phase requires significant investment in infrastructure, data collection, and specialized staff, but it offers long-term benefits in terms of highly accurate and customized AI solutions.

Overall, the framework's structured and phased approach ensures that SMEs can navigate the complexities of AI adoption effectively. By addressing the key barriers and providing practical, actionable steps, the framework empowers SMEs to harness the transformative potential of AI. This approach not only facilitates successful AI integration but also positions SMEs for sustained innovation and growth in a competitive landscape.

## 6. ACKNOWLEDGEMENTS